\documentclass[journal]{IEEEtran}
\ifCLASSINFOpdf
\else
   \usepackage[dvips]{graphicx}
\fi
\usepackage{url}
\usepackage{hyperref}
\usepackage{graphicx}
\usepackage{tabularx,lipsum,xcolor}
\usepackage{amsmath}
\usepackage{amssymb}
\usepackage{multicol, multirow, booktabs, makecell}
\usepackage{cite}
\newcolumntype{Y}{>{\centering\arraybackslash}X}

\usepackage{amsfonts}
\usepackage{color}
\usepackage{colortbl}
\usepackage{caption}
\usepackage{subcaption}
\usepackage{mathtools}
\usepackage{enumitem}
\setlist[itemize]{align=parleft,left=0pt..1.0em}
\hypersetup{
    colorlinks=true,
    linkcolor=blue,
    filecolor=blue,
    citecolor=green,
    urlcolor=blue,
}
\usepackage{algorithmic}
\usepackage[ruled,vlined,linesnumbered]{algorithm2e}
\begin{document}
\title{EEND-DEMUX: End-to-End Neural Speaker Diarization via Demultiplexed Speaker Embeddings}

\author{Sung Hwan Mun, \IEEEmembership{Student Member, IEEE}, Min Hyun Han, \IEEEmembership{Student Member, IEEE}, \\ Chanyeong Moon, \IEEEmembership{Student Member, IEEE}, and Nam Soo Kim, \IEEEmembership{Senior Member, IEEE}
\thanks{This work was supported by the COMPA grant funded by the Korea government (MSIT and Police) (No. RS2023-00235082).}
\thanks{Sung Hwan Mun, Min Hyun Han, Chanyeong Moon, and Nam Soo Kim are with Department of Electrical and Computer Engineering and Institute of New Media and Communications, Seoul National University,
Seoul 08826, South Korea. (e-mail: shmun@hi.snu.ac.kr; mhhan@hi.snu.ac.kr; cymoon@hi.snu.ac.kr; nkim@snu.ac.kr) \textit{Corresponding author: Nam Soo Kim}.}}

\markboth{Journal of \LaTeX\ Class Files, Vol. 14, No. 8, August 2015}
{Shell \MakeLowercase{\textit{et al.}}: Bare Demo of IEEEtran.cls for IEEE Journals}

\makeatletter
\patchcmd{\@maketitle}
  {\addvspace{0.5\baselineskip}\egroup}
  {\addvspace{-1.0\baselineskip}\egroup} 
  {}
  {}
\makeatother

\maketitle
 
\begin{abstract}
In recent years, there have been studies to further improve the end-to-end neural speaker diarization (EEND) systems.
This letter proposes the EEND-DEMUX model, a novel framework utilizing demultiplexed speaker embeddings.
In this work, we focus on disentangling speaker-relevant information in the latent space and then transform each separated latent variable into its corresponding speech activity.
EEND-DEMUX can directly obtain separated speaker embeddings through the demultiplexing operation in the inference phase without an external speaker diarization system, an embedding extractor, or a heuristic decoding technique.
Furthermore, we employ a multi-head cross-attention mechanism to capture the correlation between mixture and separated speaker embeddings effectively.
We formulate three loss functions based on matching, orthogonality, and sparsity constraints to learn robust demultiplexed speaker embeddings.
The experimental results on the LibriMix dataset show consistently improved performance in both a fixed and flexible number of speakers scenarios.
\end{abstract}

\vspace*{-0.3cm}

\IEEEpeerreviewmaketitle
\section{Introduction}
\IEEEPARstart{S}{peaker} diarization, often referred to as the ``\textit{who spoke when?}'' problem, is the process of estimating the speech activities of multiple speakers in an input audio stream~\cite{22Park}.
It can be employed as a pre-processing part for various speech-related tasks, \textit{e.g.,} the automatic speech recognition (ASR) or speech separation systems used in multi-party conversations~\cite{20Watanabe, 23Cornell, 21Raj}.
One prominent approach for speaker diarization is the clustering-based method, which encompasses the pipeline of several separate modules, including voice activity detection, speech segmentation, speaker representation, and clustering.
Advancements in enhanced speaker embedding~\cite{23Jung1} and innovative techniques~\cite{22Kwon, 23Heo} have greatly improved the precision of clustering-based diarization.
However, most clustering-based methods struggle with overlapped speech because they assign each time segment to a single speaker.
In addition, these approaches can be sub-optimal since each module is independently optimized. It further necessitates careful engineering labor to maintain the overall pipeline~\cite{22Raj}.

To address these limitations, the end-to-end neural speaker diarization framework (EEND) was introduced~\cite{19Fujita1, 19Fujita2}. This framework directly predicts the speech activities of multiple speakers from an input mixture. Fujita \textit{et al.}~\cite{19Fujita1} reformulated speaker diarization as a frame-wise multi-label classification problem and optimized an integrated network by directly minimizing diarization errors using the permutation invariant training (PIT) loss function~\cite{17Yu}.
EEND, while effective, has a limitation with its fixed number of output heads, making it less flexible when dealing with scenarios involving unknown number of speakers. 
To mitigate this, Horiguchi \textit{et al.}~\cite{20Horiguchi, 22Horiguchi} introduced EEND-EDA, which employed a sequence-to-sequence method with an LSTM encoder-decoder module to calculate speaker-wise attractors from frame-wise embeddings and predict the speaker existence probabilities from the obtained attractors.
This dynamic approach enables the system to accommodate varying numbers of speakers.

Recent research efforts have sought to enhance the performance of EEND system further, such as regularization techniques for multi-head self-attention in Transformer encoder blocks~\cite{22Yu, 23Jeoung1, 23Jeoung2}, structural design for attractors~\cite{22Rybicka, 23Fujita, 23Broughton}, utilization of speaker-specific prior information~\cite{20Medennikov, 20Medennikov2, 23Wang1, 23Boeddeker, 23Chen, 23Chen2, 23Wang2}, and multi-task learning approaches~\cite{21Raj, 23Maiti, 23Ao}. 
Particularly, multi-task learning methods jointly trained speaker diarization and speech separation tasks within a single neural network, demonstrating improved performance due to the complementarity of both tasks.
Although these approaches showed that joint modeling of two tasks could provide a synergistic effect to each other towards mutual benefits, when focusing on our primary objective, speaker diarization, concurrently solving both optimization problems for two tasks may increase system complexity, potentially leading to sub-optimal performance.

In this letter, we propose an EEND framework based on \textit{demultiplexed speaker embeddings} (EEND-DEMUX).
Rather than modeling entire speaker diarization and speech separation networks, we focus on disentangling speaker-relevant information in the latent space and subsequently transform each separated latent variable to its corresponding speech activity.
Unlike previous approaches~\cite{20Medennikov, 20Medennikov2, 23Chen, 23Chen2}, requiring an external speaker diarization system, an embedding extractor, or a heuristic decoding technique during the inference phase, EEND-DEMUX can directly obtain separated speaker embeddings through the demultiplexing operation, which can promote a more efficient speaker diarization process.
We also define objective functions to effectively prompt the demultiplexing process over the speaker embeddings.
The main contributions of this work can be summarized as follows:
\begin{itemize}
    \item We incorporate a frame-wise demultiplexing process in the speaker embedding space to the EEND framework.
    \item To facilitate the learning of demultiplexed speaker embeddings, we define the objective functions that adhere to three constraints: matching, orthogonality, and sparsity.
    \item We also leverage a multi-head cross-attention mechanism (MHCA) to capture the correlation between mixture and separated speaker embeddings effectively.
    \item Experimental results using 2-speaker and 3-speaker datasets demonstrated improved performance in scenarios involving both fixed and flexible numbers of speakers.
\end{itemize}

\vspace{-0.35cm}
\section{Proposed EEND-DEMUX Framework}
\label{sec2}
\subsection{Model Architecture}
\begin{figure}[t]
    \centering
    \includegraphics[width=0.90\linewidth]{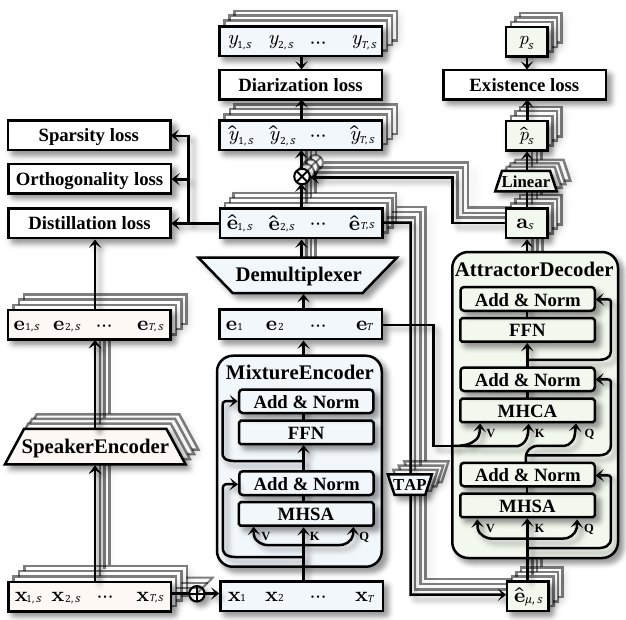}
    \caption{The overview of the EEND-DEMUX model.}
    \label{fig:framework}
\end{figure}

Let $\mathbf{x}_{t}\in \mathbb{R}^{F}$ be an $F$-dimensional input speech feature at the $t$-th frame where $t\in\{1,\dots,T\}$.
We assume that an input speech feature $\mathbf{x}_{t}$ is observed as a sum of $S$ speaker speech features as given by:
\begin{gather}
    \mathbf{x}_{t} = \sum\nolimits_{s=1}^{S} y_{t,s}\,\mathbf{x}_{t,s} \in \mathbb{R}^{F},
\end{gather}
where $\mathbf{x}_{t,s}\in \mathbb{R}^{F}$ and $y_{t,s} \in \{0,1\}$ respectively represent an $F$-dimensional source speech feature and a speech activity label for speaker $s$ at the $t$-th frame.
As illustrated in Fig. \ref{fig:framework}, the EEND-DEMUX framework comprises four modules: MixtureEncoder, Demultiplexer, AttractorDecoder, and SpeakerEncoder.
First, the MixtureEncoder module takes a sequence of input mixture features and outputs a sequence of mixture embeddings as follows:
\begin{gather}
  \mathbf{E} = \text{MixtureEncoder}(\mathbf{X}) \in \mathbb{R}^{D\times T},
\end{gather}
where $\mathbf{X}=[\mathbf{x}_{1},\dots,\mathbf{x}_{T}]\in\mathbb{R}^{F\times T}$ and $\mathbf{E}=[\mathbf{e}_{1},\dots,\mathbf{e}_{T}]\in\mathbb{R}^{D\times T}$ respectively represent a $T$-length sequence of $F$-dimensional input mixture features and a $T$-length sequence of mixture embeddings.
MixtureEncoder$:\mathbb{R}^{F\times T}\mapsto \mathbb{R}^{D\times T}$ consists of stacked Transformer encoder blocks.
Next, a mixture embedding $\mathbf{e}_{t}$ is separated into $S$ speaker embeddings $[\hat{\mathbf{e}}_{t,1},\dots,\hat{\mathbf{e}}_{t,S}]$ at every frame by the Demultiplexer. 
Specifically, a $T$-length sequence of mixture embeddings $\mathbf{E}$ is fed into the Demultiplexer module as:
\begin{gather}
  [\hat{\mathbf{E}}_{1},\dots,\hat{\mathbf{E}}_{S}] = \text{Demultiplexer}(\mathbf{E}) \in \mathbb{R}^{D\times T \times S}.
\end{gather}
Here, $\hat{\mathbf{E}}_{s}=[\hat{\mathbf{e}}_{1,s},\dots,\hat{\mathbf{e}}_{T,s}]\in\mathbb{R}^{D\times T}$ represents a $T$-length sequence of $D$-dimensional embeddings for speaker $s$
where we refer to each $\hat{\mathbf{e}}_{t,s}$ as a \textit{demultiplexed speaker embedding}.

After that, $D$-dimensional $S$ attractors $[\mathbf{a}_{1},\dots,\mathbf{a}_{S}]$ are extracted by the AttractorDecoder as follows:
\begin{gather}
  \mathbf{A} = \text{AttractorDecoder}([\hat{\mathbf{e}}_{\mu,1},\dots,\hat{\mathbf{e}}_{\mu,S}],\>\mathbf{E}) \in \mathbb{R}^{D\times S},
\end{gather}
where $\mathbf{A}=[\mathbf{a}_{1},\dots,\mathbf{a}_{S}]\in\mathbb{R}^{D\times S}$ indicates a collection of $S$ attractors.
Also, $[\hat{\mathbf{e}}_{\mu,1},\dots,\hat{\mathbf{e}}_{\mu,S}]\in\mathbb{R}^{D\times S}$ represents a collection of $D$-dimensional $S$ prototype speaker embeddings where $\hat{\mathbf{e}}_{\mu,s} = {1 \over T} \sum\nolimits_{t} \hat{\mathbf{e}}_{t,s} \in \mathbb{R}^{D}$ denotes the prototype speaker embedding for speaker $s$ obtained through temporal average pooling (TAP).
Analogous to~\cite{22Rybicka, 23Fujita, 23Chen, 23Chen2}, the AttractorDecoder module includes both a multi-head self-attention (MHSA) block and a multi-head cross-attention (MHCA) block.
The MHSA block facilitates the interaction between $S$ prototype speaker embeddings $[\hat{\mathbf{e}}_{\mu,1},\dots,\hat{\mathbf{e}}_{\mu,S}]$, leading to more distinct attractors.
Meanwhile, the MHCA block enables $S$ prototype speaker embeddings $[\hat{\mathbf{e}}_{\mu,1},\dots,\hat{\mathbf{e}}_{\mu,S}]$ to concentrate on $T$-length mixture embeddings $[\mathbf{e}_{1},\dots,\mathbf{e}_{T}]$, enhancing the ability to capture the correlation between each other.

Finally, a posterior probability $\hat{y}_{t,s}$ at the $t$-th frame for speaker $s$ is estimated by applying a sigmoid activation function to a dot product between the demultiplexed speaker embedding $\hat{\mathbf{e}}_{t,s}$ and the attractor $\mathbf{a}_{s}$:
\begin{gather}
    \hat{y}_{t,s} = \mathrm{sigmoid} (\hat{\mathbf{e}}_{t,s}^{\top} \, \mathbf{a}_{s}) \in (0,1),
\end{gather}
where $\mathrm{sigmoid(\cdot)}$ denotes the sigmoid activation function.

Additionally, a speaker existence probability $\hat{p}_{s}$ for speaker $s$ is obtained from the attractor $\mathbf{a}_{s}$ as follows:
\begin{gather}
    \hat{p}_{s} = \mathrm{sigmoid}(\mathrm{Linear}(\mathbf{a}_{s})) \in (0,1),
\end{gather}
where $\mathrm{Linear}(\cdot):\mathbb{R}^{D}\mapsto \mathbb{R}$ represents a fully connected layer.
As the EEND-DEMUX architecture yields the fixed $S$ posterior probabilities,
to deal with varying numbers of speakers, we take speaker-active components among the $S$ output heads by defining the valid speaker set $\mathcal{S}$ as:
\begin{gather}
    \mathcal{S} = \{ s \> | \> s \in \{1,\dots,S\} \> \wedge \> p_{s} \geq 0.5 \},
\end{gather}
where $p_{s}\in\{0,1\}$ is a ground-truth speaker existence label.
During the inference phase, the valid speaker set $\mathcal{S}$ is determined using the predicted speaker existence probability $\hat{p}_{s}$ instead of the ground-truth label $p_{s}$.

Furthermore, during training, the SpeakerEncoder module estimates a sequence of oracle speaker embeddings by taking a sequence of source speech features as follows:
\begin{gather}
    \mathbf{E}_{s} = \text{SpeakerEncoder}(\mathbf{X}_{s}) \in \mathbb{R}^{D\times T}.
\end{gather}
Here, $\mathbf{E}_{s}=[\mathbf{e}_{1,s},\dots,\mathbf{e}_{T,s}]$ represents a $T$-length sequence of $D$-dimensional oracle speaker embeddings for speaker $s$.
To obtain a reliable speaker representation, we pre-train it with the speaker classification loss function using the source speech features and the corresponding speaker identities.

\vspace{-0.3cm}
\subsection{Training Objective Functions}
The objective function used to train the proposed model consists of two losses: \textit{speaker diarization} and \textit{demultiplexing} losses.
The speaker diarization loss includes the objectives used in the EEND-EDA framework~\cite{20Horiguchi, 22Horiguchi}, \textit{i.e.}, the diarization loss and the existence loss.
First, the diarization loss is defined as the binary cross-entropy between the estimated posterior $\hat{y}_{t,s}$ and the target label $y_{t,s}$ as follows:
\begin{gather}
    \label{loss:diar}
    \mathcal{L}_{\textnormal{diar}} = {1 \over T|\mathcal{S}|} \min_{\phi \in \Phi(\mathcal{S})} \sum\nolimits_{t=1}^{T}\sum\nolimits_{s\in\mathcal{S}} H(y_{t,s}^{\phi}, \hat{y}_{t,s}),
\end{gather}
where $H(y_{t,s}, \hat{y}_{t,s}) := -y_{t,s}\log \hat{y}_{t,s} - (1-y_{t,s}) \log (1-\hat{y}_{t,s})$ and $|\mathcal{S}|$ indicates the number of elements in the valid speaker set $\mathcal{S}$.
Further, $\Phi(\mathcal{S})$ denotes the set of all possible permutations of $\{1,\dots,|\mathcal{S}|\}$ with $\phi=(\phi_{1},\dots,\phi_{|\mathcal{S}|})$ representing a permutation and $y^{\phi}_{t,s}$ is the speech activity at the $t$-th frame for speaker $s$ under permutation $\phi$.
The existence loss is calculated by the binary cross-entropy between the predicted speaker existence probability $\hat{p}_{s}$ and the ground-truth label $p_{s}$:
\begin{gather}
    \label{loss:ext}
    \mathcal{L}_{\textnormal{ext}} = {1 \over S} \sum\nolimits_{s=1}^{S} H(p_{s}, \hat{p}_{s}).
\end{gather}

The demultiplexing loss aims to disentangle demultiplexed speaker embeddings $\{\hat{\mathbf{e}}_{t,s}|{s\in\mathcal{S}}\}$ from a mixture embedding $\mathbf{e}_{t}$ at the frame level.
To facilitate the disentangling process over the demultiplxed speaker embeddings, we define three objective functions based on the following constraints:
\begin{itemize}
    \item Matching constraint: Each demultiplexed speaker embedding is enforced to match the oracle speaker embedding extracted from its source speech feature.
    \item Orthogonality constraint: $|\mathcal{S}|$ demultiplexed speaker embeddings separated from a mixture embedding should be orthogonal to each other on the speaker basis in every frame.
    \item Sparsity constraint: All separated $D$-dimensional demultiplexed speaker embeddings should be sparse, \textit{i.e.}, having minimum non-zero elements.
\end{itemize}
For the matching constraint, we adopt the distillation strategy to transfer the knowledge from the oracle speaker embedding $\mathbf{e}_{t,s}$ to the demultiplexed speaker embedding $\hat{\mathbf{e}}_{t,s}$ at the frame level.
Given an optimal permuted sequence $\phi^{*}$ obtained from (\ref{loss:diar}), the distillation loss is defined as follows:
\begin{gather}
    \label{loss:dis}
    \mathcal{L}_{\textnormal{dis}} = {1 \over T|\mathcal{S}|} \sum\nolimits_{t=1}^{T}\sum\nolimits_{s\in\mathcal{S}} || \mathbf{e}_{t,s}^{\phi^*} - \hat{\mathbf{e}}_{t,s}^{\phi^*} ||_{2},
\end{gather}
where $\mathbf{e}_{t,s}^{\phi^*}$ and $\hat{\mathbf{e}}_{t,s}^{\phi^*}$ respectively represent the oracle and demultiplexed speaker embeddings for speaker $s$ at the $t$-th frame aligned with the optimal permutation $\phi^{*}$.

As for the orthogonality constraint, we enforce constraints to ensure that the demultiplexed speaker embeddings from different speakers are orthogonal to each other, while those from the same speaker are similar.
Motivated by~\cite{21Ranasinghe}, we define the orthogonality loss that simultaneously promotes both inter-speaker orthogonality and intra-speaker clustering through the cosine similarity within a mini-batch:
\begin{gather}
    \label{loss:ort}
    \mathcal{L}_{\textnormal{ort}} = {1 \over T {|\mathcal{S}| \choose 2} } \sum_{t=1 \>}^{T} \, \sum_{\mathclap{\substack{\>\>\>(i,j)\in\mathcal{S}\\ \>\> 1 \leq i < j \leq |\mathcal{S}|}}} \>\>\> \underbrace{1 - \langle \hat{\mathbf{e}}_{t,i}^{\phi^*}, \hat{\mathbf{e}}_{\mu,i}^{\phi^*}\rangle}_{\>\>\>\>\>\>\> \mathrm{positive \, term}} + \underbrace{|\langle \hat{\mathbf{e}}_{t,i}^{\phi^*}, \hat{\mathbf{e}}_{t,j}^{\phi^*}\rangle|}_{\mathrm{negative \, term}}, 
\end{gather}
where the operator $\langle \cdot\,,\cdot \rangle$ indicates the cosine similarity between two vectors.
For given the demultiplexed speaker embedding $\hat{\mathbf{e}}_{t,i}$ for speaker $i$ at the $t$-th frame,
its prototype speaker embedding $\hat{\mathbf{e}}_{\mu,i}$ serves as a positive sample and the demultiplexed speaker embedding $\hat{\mathbf{e}}_{t,j}$ from different speaker $j$ at the $t$-th frame is employed as a negative sample.

On top of that, we impose a structural constraint to encourage each demultiplexed speaker embedding to be sparse.
In previous studies~\cite{15Faruqui, 16Bhowmik, 21Medini}, sparsifying dense embeddings has enhanced the interpretation ability for learning representations and also alleviated over-fitting, which could shrink the complexity induced by the models and improve the generalization performance on unseen data.
To learn effectively separated demultiplexed speaker embeddings, we apply a loss that regularizes the sparsity by penalizing the $\ell_1$-norm of all the demultiplexed speaker embeddings $\{\hat{\mathbf{x}}_{t,s}|\, \forall \>t,s\}$ as follows:
\begin{gather}
    \label{loss:spa}
    \mathcal{L}_{\textnormal{spa}} = {1 \over T|\mathcal{S}|} \sum\nolimits_{t=1}^{T}\sum\nolimits_{s\in\mathcal{S}} || \hat{\mathbf{e}}_{t,s} ||_{1}.
\end{gather}
Since the structural sparseness of demultiplexed speaker embeddings obtained by regularizing the loss (\ref{loss:spa}) can suppress the model complexity, we expect the more advanced generalization ability and the more enhanced separability in the demultiplexing process.
Finally, combining the speaker diarization and the demultiplexing objective functions, we construct the total objective function $\mathcal{L}_{\textnormal{total}}$ as given by:
\begin{equation}
\begin{split} 
    \label{loss:total}
    \mathcal{L}_{\textnormal{total}} = \lambda_{\textnormal{diar}} \mathcal{L}_{\textnormal{diar}} & + \lambda_{\textnormal{ext}} \mathcal{L}_{\textnormal{ext}} \\
                  + \> & \lambda_{\textnormal{dis}} \mathcal{L}_{\textnormal{dis}} + \lambda_{\textnormal{ort}} \mathcal{L}_{\textnormal{ort}} + \lambda_{\textnormal{spa}} \mathcal{L}_{\textnormal{spa}},    
\end{split}
\end{equation} 
where $\lambda_{\textnormal{diar}}$, $\lambda_{\textnormal{ext}}$, $\lambda_{\textnormal{dis}}$, $\lambda_{\textnormal{ort}}$, and $\lambda_{\textnormal{spa}}$ are weighting factors for balancing each objective function.

\vspace{-0.25cm}
\section{Experiments}
\subsection{Training and Test Datasets}
We evaluated the performance of the proposed technique using the LibriMix~\cite{20Cosentino} dataset, which includes training and test mixtures generated by samples from LibriSpeech~\cite{15Panayotov} train-clean100 and test-clean with WHAM!~\cite{19Wichern} at a 16kHz sampling rate.
The dataset comprises two-speaker mixtures (Libri2Mix) and three-speaker mixtures (Libri3Mix), which include 58h/11h and 40h/11h mixtures of training/test sets, respectively.
For experiments on the scenario of a flexible number of speakers, we employed the dataset combining both Libri2Mix and Libri3Mix (Libri2\&3Mix).
During training, we used the \textit{min} mode to compare with previous works~\cite{23Maiti, 23Ao}.

\vspace{-0.35cm}
\subsection{Configurations}
We used 80-dimensional log-mel filterbanks (LMF) as acoustic features with a window size of 25 ms and a frame-shift of 10 ms for all networks.
The dimension of the embeddings was set to 256.
MixtureEncoder is implemented using 4-stacked Transformer encoders with four attention heads.
Demultiplexer consists of $S$ parallel 2-stacked 1D-CNN modules where each module sequentially includes a CNN layer with a kernel size of 5, a batch normalization, and a ReLU activation function. AttractorDecoder is designed using 2-stacked Transformer decoders with four attention heads without a positional encoding.
SpeakerEncoder utilizes the the selective kernel attention-based 1D-CNN network (SKA-TDNN)~\cite{23Mun}. 
For pre-training, we first trained the network with VoxCeleb datasets~\cite{17Nagrani, 18Chung} and then fine-tuned it using the source speech feature of the LibriMix train set with the target speaker identity.
After pre-training, we froze the network parameters and employed the last frame-level output feature before the pooling layer as the oracle speaker embedding.

\vspace{-0.35cm}
\subsection{Implementation Details}
We conducted experiments using PyTorch and optimized the model via Adam using a global batch size of 128 with accumulated gradients of 2 on three parallel NVIDIA GeForce RTX 3090 GPUs. We utilized the Noam scheduler with a maximum learning rate of $5\textnormal{e-}4$ and a warm-up of 30 epochs.
The weights for total objective function are set to $\lambda_{\textnormal{diar}}=1$, $\lambda_{\textnormal{ext}}=1\textnormal{e-}2$, $\lambda_{\textnormal{dis}}=2.5$, $\lambda_{\textnormal{ort}}=1\textnormal{e-}3$, and $\lambda_{\textnormal{spa}}=1\textnormal{e-}5$.

\vspace{-0.35cm}
\subsection{Evaluation Metric}
For the evaluation metric, we reported the diarization error rate (DER, \%), which includes false alarm (FA, \%), missed detection (MS, \%), and speaker confusion (SF, \%); the lower, the better for all.
The collar tolerance is set to 0 sec.

\begin{table}[t]
  \caption{Experimental results on a fixed 2-speaker scenario (Libri2Mix) for \textit{min} mode in terms of DER, FA, and MI (\%). $\dagger$ denotes our re-implementation.}
   \centering
  \label{tab:fix_2mix}
  \vspace{-4.0pt}
\scriptsize{
\begin{tabular}{l cccc}
\toprule
Method &  \multicolumn{1}{c}{DER}    & \multicolumn{1}{c}{FA}      & \multicolumn{1}{c}{MI}      & \multicolumn{1}{c}{CF} \\

\midrule
SA-EEND$^{\dagger}$           & 6.13    & 3.54    & 2.11    & 0.48 \\
EEND-EDA~\cite{23Maiti}       & 5.93    & -       & -       & -    \\

\midrule
EEND-SS ($\mathcal{L}_{\text{diar}}+\mathcal{L}_{\text{ext}}$)~\cite{23Maiti}         & 5.26          & -             & -             & -             \\
 $+\,\mathcal{L}_{\text{SI-SDR}}$                                                     & 5.12          & -             & -             & -             \\
 $+\,\mathcal{L}_{\text{SI-SDR}}+\mathcal{L}_{\text{SI-SDR}}\,+\,$LMF                 & 5.02          & -             & -             & -             \\

\midrule
EEND-DEMUX ($\mathcal{L}_{\text{diar}}+\mathcal{L}_{\text{ext}}$)                     & 4.67          & 2.83          & 1.52          & 0.32          \\
 $+\,\mathcal{L}_{\text{dis}}$                                                        & 3.94          & 2.45          & 1.29          & 0.20          \\
 $+\,\mathcal{L}_{\text{dis}}+\mathcal{L}_{\text{ort}}$                               & 3.85          & 2.41          & 1.26          & 0.18          \\
 $+\,\mathcal{L}_{\text{dis}}+\mathcal{L}_{\text{ort}}+\mathcal{L}_{\text{spa}}$      & \textbf{3.79} & \textbf{2.38} & \textbf{1.25} & \textbf{0.16} \\

\bottomrule

\end{tabular}
}
\end{table}
\begin{table}[t]
  \caption{Experimental results on a fixed 3-speaker scenario (Libri3Mix) for \textit{min} mode.}
   \centering
  \label{tab:fix_3mix}
  \vspace{-4.0pt}
\scriptsize{
\begin{tabular}{l cccc}
\toprule
Method &  \multicolumn{1}{c}{DER}    & \multicolumn{1}{c}{FA}      & \multicolumn{1}{c}{MI}      & \multicolumn{1}{c}{CF} \\

\midrule
SA-EEND$^{\dagger}$          & 9.05    & 5.92    & 2.72    & 0.41 \\
EEND-EDA~\cite{23Maiti}      & 8.81    & -       & -       & -    \\

\midrule
EEND-SS ($\mathcal{L}_{\text{diar}}+\mathcal{L}_{\text{ext}}$)~\cite{23Maiti}        & 6.50          & -             & -             & -             \\
 $+\,\mathcal{L}_{\text{SI-SDR}}$                                                    & 6.26          & -             & -             & -             \\
 $+\,\mathcal{L}_{\text{SI-SDR}}+\mathcal{L}_{\text{SI-SDR}}\,+\,$LMF                & 6.00          & -             & -             & -             \\

\midrule
EEND-DEMUX ($\mathcal{L}_{\text{diar}}+\mathcal{L}_{\text{ext}}$)                    & 5.64          & 3.51          & 1.68          & 0.45          \\
 $+\,\mathcal{L}_{\text{dis}}$                                                       & 5.13          & 3.23          & 1.51          & 0.39          \\
 $+\,\mathcal{L}_{\text{dis}}+\mathcal{L}_{\text{ort}}$                              & 4.96          & 3.14          & 1.44          & 0.38          \\
 $+\,\mathcal{L}_{\text{dis}}+\mathcal{L}_{\text{ort}}+\mathcal{L}_{\text{spa}}$     & \textbf{4.91} & \textbf{3.12} & \textbf{1.42} & \textbf{0.37} \\

\bottomrule

\end{tabular}
  }
\end{table}

\vspace{-0.35cm}
\subsection{Results on the Fixed Number of Speakers}
We conducted experiments on the scenario for a fixed number of speakers.
The proposed models were evaluated on 2-speaker and 3-speaker conditions using the test sets \textit{min} mode of Libri2Mix and Libri3Mix, respectively.
For the 2-speakers scenario, the model was trained using the training set \textit{min} mode of Libri2Mix.
For the training of the 3-speakers scenario, similar to~\cite{20Horiguchi, 23Maiti}, we had the adaptation strategy; the weights pre-trained with the training set of Libri2Mix were initialized, and then the model for the 3-speaker scenario was fine-tuned with the training set of Libri3Mix.
We compared our models with the state-of-the-art EEND models on the LibriMix dataset reported in~\cite{23Maiti, 23Ao}.
We also reported the results of EEND-DEMUX models trained using the four combinations of proposed losses, 
\textit{i.e.,} (1) only speaker diarization (SD) loss, (2) the SD and distillation losses, (3) the SD, distillation, and orthogonality losses, and (4) the total objective loss.
As shown in both Tables~\ref{tab:fix_2mix} and~\ref{tab:fix_3mix}, imposing the demultiplexing constraint terms (the distillation, orthogonality, and sparsity) consistently improves the diarization performance in all metrics.
Especially, the best-performing model achieves DERs of 3.79\% and 4.91\% for 2-speaker and 3-speaker scenarios, respectively.
These results outperform the best baseline models with relative improvements of 24.50\% and 18.17\%, respectively.

\begin{table}[t]
  \caption{Experimental results on flexible number of speakers scenario (Libri2\&3Mix) for \textit{min} mode.}
   \centering
  \label{tab:flex_min}
  \vspace{-4.0pt}
\scriptsize{
\begin{tabular}{l cccc}
\toprule
Method &  \multicolumn{1}{c}{DER}    & \multicolumn{1}{c}{FA}      & \multicolumn{1}{c}{MI}      & \multicolumn{1}{c}{CF} \\

\midrule
SA-EEND$^{\dagger}$     & 10.38   & 5.20    & 4.76    & 0.42 \\
EEND-EDA~\cite{23Maiti} & 10.16   & -       & -       & -    \\
TS-VAD~\cite{23Ao}      & 5.30    & 2.83    & 2.32    & 0.15 \\

\midrule
EEND-SS ($\mathcal{L}_{\text{diar}}+\mathcal{L}_{\text{ext}}$)~\cite{23Maiti}       & 8.79          & -             & -             & -   \\
 $+\,\mathcal{L}_{\text{SI-SDR}}$                                                   & 6.27          & -             & -             & -   \\
 $+\,\mathcal{L}_{\text{SI-SDR}}+\mathcal{L}_{\text{SI-SDR}}\,+\,$LMF               & 6.04          & -             & -             & -   \\

\midrule
USED ($\mathcal{L}_{\text{diar}}+\mathcal{L}_{\text{ext}}$)~\cite{23Ao}             & 4.72          & 2.80          & 1.83          & 0.10 \\
 $+\,\mathcal{L}_{\text{spk}}$                                                      & 4.48          & 2.60          & 1.79          & 0.08 \\

\midrule
EEND-DEMUX ($\mathcal{L}_{\text{diar}}+\mathcal{L}_{\text{ext}}$)                   & 4.96          & 2.95          & 1.89          & 0.15            \\
 $+\,\mathcal{L}_{\text{dis}}$                                                      & 4.55          & 2.61          & 1.82          & 0.12            \\
 $+\,\mathcal{L}_{\text{dis}}+\mathcal{L}_{\text{ort}}$                             & 4.44          & 2.57          & 1.78          & 0.09            \\
 $+\,\mathcal{L}_{\text{dis}}+\mathcal{L}_{\text{ort}}+\mathcal{L}_{\text{spa}}$    & \textbf{4.39} & \textbf{2.54} & \textbf{1.77} & \textbf{0.08}   \\

\bottomrule
\end{tabular}
  }
\end{table}

\vspace{-0.35cm}
\subsection{Results on the Flexible Number of Speakers}
To evaluate our model on the scenario for a flexible number of speakers, we combined Libri2Mix and Libri3Mix datasets (Libri2\&3Mix).
For the training, we had an adaptation step with pre-trained weights of the \textit{min} mode of Libri2Mix and fine-tuned the model using the combined training set \textit{min} mode of Libri2\&3Mix.
Tables~\ref{tab:flex_min} and~\ref{tab:flex_max} show the evaluation results on the \textit{min} and \textit{max} modes for the combined test set, Libri2\&3Mix.
Using the proposed objective functions leads to the improvement in diarization performance.
The best performance is a DER of 4.39\% for \textit{min} mode and a DER of 5.08\% for \textit{max} mode, outperforming the state-of-the-art baselines with relative improvements of 2.00\% and 3.05\%.

\begin{table}[t]
  \caption{Experimental results on flexible number of speakers scenario (Libri2\&3Mix) for \textit{max} mode.}
   \centering
  \label{tab:flex_max}
  \vspace{-4.0pt}
\scriptsize{
\begin{tabular}{l cccc}
\toprule
Method &  \multicolumn{1}{c}{DER}    & \multicolumn{1}{c}{FA}      & \multicolumn{1}{c}{MI}      & \multicolumn{1}{c}{CF} \\

\midrule
SA-EEND$^{\dagger}$                   & 11.42   & 6.21    & 4.50    & 0.71 \\
wav2vec2.0 \textsc{Base}~\cite{23Ao}  & 7.62    & 2.28    & 4.82    & 0.52 \\ 
HuBERT \textsc{Base}~\cite{23Ao}      & 7.56    & 2.40    & 4.81    & 0.35 \\
TS-VAD~\cite{23Ao}                    & 7.28    & 2.78    & 3.61    & 0.89 \\

\midrule
USED ($\mathcal{L}_{\text{diar}}+\mathcal{L}_{\text{ext}}$)~\cite{23Ao} \>\>\>\>\>\,\,    & 6.49          & 2.81          & 3.04          & 0.65      \\
 $+\,\mathcal{L}_{\text{spk}}$                                                            & 5.24          & 2.55          & 2.42          & 0.27      \\

\midrule
EEND-DEMUX ($\mathcal{L}_{\text{diar}}+\mathcal{L}_{\text{ext}}$)                         & 6.81          & 3.05          & 3.12          & 0.64           \\
 $+\,\mathcal{L}_{\text{dis}}$                                                            & 5.19          & 2.50          & 2.44          & 0.25           \\
 $+\,\mathcal{L}_{\text{dis}}+\mathcal{L}_{\text{ort}}$                                   & 5.12          & 2.49          & 2.41          & \textbf{0.22}  \\
 $+\,\mathcal{L}_{\text{dis}}+\mathcal{L}_{\text{ort}}+\mathcal{L}_{\text{spa}}$          & \textbf{5.08} & \textbf{2.46} & \textbf{2.38} & 0.24           \\

\bottomrule
\end{tabular}
  }
\end{table}

\section{Conclusion}
This letter proposed the EEND-DEMUX model, a novel end-to-end neural speaker diarization framework via demultiplexed speaker embeddings.
We incorporated a frame-wise demultiplexing process into the EEND framework and defined the objective functions to learn demultiplexed speaker embeddings effectively.
The experimental results on the LibriMix dataset showed consistently improved performance in both a fixed and flexible number of speaker scenarios.
We plan to pursue future work to utilize this framework as a key component in different tasks, such as multi-speaker automatic speech recognition or speech separation tasks.

\end{document}